\documentclass[%
 reprint,
 amsmath,amssymb,
 aps,
prb,
]{revtex4-2}

\usepackage{graphicx}
\usepackage{dcolumn}
\usepackage{bm}
\usepackage{xcolor}

\DeclareMathOperator{\sign}{sign}       

\usepackage{hyperref}

\begin{document}

\preprint{APS/123-QED}

\title{Coexistence of gapless and gapped vortex modes \\ with Majorana corner states in a 2D second-order topological superconductor}

\author{A.\,D.\, Fedoseev}
 \email{fad@iph.krasn.ru}
\author{A.\,O.\, Zlotnikov}%
 \email{zlotn@iph.krasn.ru}

	\affiliation{%
		Kirensky Institute of Physics, Federal Research Center KSC SB RAS, 660036 Krasnoyarsk, Russia}

\date{\today}

\begin{abstract}
Although the appearance of vortex-localized states with zero energy in first-order topological superconductors is well known, their possibility to form in the higher-order topological phase of 2D systems has not been completely uncovered yet. Here we demonstrate the coexistence of zero-energy vortex modes and Majorana corner modes in the model of a 2D second-order topological superconductor. The model describes an interface between a normal layer supporting the topological insulating phase and a superconducting layer, for which different symmetries of the spin-singlet superconducting order parameter are considered. We show that the gapless vortex modes can appear under certain conditions in the superconducting state with a vortex if the bulk energy spectrum of the normal (non-superconducting) state is gapless and has Dirac cones. The number of pairs of such vortex modes corresponds to the number of Dirac cones. It is essential that if the normal bulk spectrum becomes gapped and the system is in the state of a topological insulator, then the zero-energy vortex modes can not be realized, while Majorana corner modes hold in the superconducting state. The interaction of the vortex modes with the edge and topological corner modes is studied when the vortex appears near the boundaries.
\end{abstract}

\maketitle


\section{\label{sec1}Introduction}

The development of the concept of topologically nontrivial systems has led in recent years to an active study of higher-order topological insulators and superconductors (HOTSCs) \cite{benalcazar-17,Langbehn-17,Zhu-18,Liu-18,WangY-18,Yan2018}. The spectrum of their both bulk and edge states has a gap. In turn, topologically protected gapless excitations arise, being localized at the boundaries of higher orders, i.e. at corners (corners and hinges) in 2D (3D) systems \cite{volovik-10}. It is important to note that in the case of 2D HOTSCs such states are Majorana corner modes (MCMs) which possess zero energy and can be used for the braiding procedure \cite{Pahomi2020,Zhang2020PRR,Zhang2020PRB} as obey non-Abelian exchange statistics \cite{ivanov-01,alicea-11}. These features are promising for the realization of quantum computation \cite{Nayak2008}.

It is well known that Majorana zero modes formed near the boundaries and Majorana zero modes localized at the vortex cores correspond to each other in first-order topological superconductors (FOTSCs)~\cite{alicea-12}. It is supposed that the existence of one type of such excitations (edge or vortex) guarantees the presence of excitations of the other type in the same system. The interplay between Majorana edge and vortex states has been extensively studied in different FOTSCs, such as triplet $p_x + ip_y$-wave superconductors \cite{stern-04, gurarie-07}, superconducting structures with spin-orbit interaction~\cite{sau-10, bjornson-13}, topological insulator and superconductor heterostructures~\cite{fu-08, rakhmanov-11, chiu-11}, superconducting doped topological insulators~\cite{hosur-11}, spin-singlet superconductors with noncollinear spin ordering~\cite{zlotnikov-23} and others. The vector potential of the magnetic field modify the interaction between vortex and edge modes, since the second Majorana mode (called ``exterior'') is localized in this case on the characteristic magnetic length from the vortex core hosting the first Majorana mode~\cite{akzyanov-16}. Therefore, if the magnetic length is less than the system size, then the vortex and edge states do not interact with each other. Otherwise, if the magnetic length exceeds the system size, the second Majorana mode is localized on the boundaries.

Nevertheless, the concept of such edge-vortex correspondence breaks in HOTSCs. Since the edge spectrum of HOTSC must be gapped, it is natural to expect the gapped excitations of vortex states from the point of view of FOTSC. Yet, the coexistence of Majorana hinges modes and zero modes localized on the opposite ends of a vortex line in 3D HOTSC is shown in~\cite{kheirkhah-21}, as well as vortex and corner modes have been found simultaneously in a system with the space group $P4/nmm$ in the presence of superconductivity and antiferromagnetic ordering~\cite{zhang-24}. Moreover, it has been proposed that the bulk-vortex correspondence should be considered to predict the existence of zero-energy vortex modes in HOTSCs~\cite{zhang-22}.

It should be noted that there are proposals for realization of HOTSC state in FeTe$_{0.55}$Se$_{0.45}$ \cite{zhang-19, zhang-21}, that is usually considered as FOTSC \cite{wang-18, pathak-21, machida-23}. There is the experimental evidence that the helical hinge modes are realized in FeTe$_{0.55}$Se$_{0.45}$ \cite{gray-19}. Moreover, this compound is an ideal platform to experimentally study the zero-energy vortex modes \cite{machida-23}, since the ratio of the square of the superconducting gap and Fermi energy, which characterize the energy of vortex-localized Caroli–de Gennes–Matricon states, is high (an order of 100 $\mu\text{eV}$) enough to resolve zero-energy modes by STM. Thus, the problem of coexistence of vortex modes and corner or hinge modes in HOTSCs might also be considered experimentally.

In the present study we describe the interplay of the gapped and gapless vortex modes and Majorana corner states in the 2D model of HOTSC describing an interface between a two-band normal layer with spin-orbit interaction and a superconducting layer. Usually, a normal layer is considered as a topological insulator with the gapped bulk spectrum and gapless edge spectrum~\cite{wangQ-18, aksenov-23}. Superconducting pairings can induce a Dirac mass in the edge spectrum and support the formation of HOTSC state~\cite{aksenov-23-2}. As it will be shown in the present study, the vortex modes are also gapped in such regime. Therefore, we focus on the case, when the normal layer has one or two Dirac cones in the bulk spectrum. In this case superconducting pairings induce a gap in the bulk energy spectrum, as well as in the edge spectrum, and also lead to formation of HOTSC state. The difference from the previous case here is that such regime supports the single-pair or double-pair vortex-localized modes with zero energy which coexist with Majorana corner modes. For the double-pair vortex modes the vortex core hosts two pairs of zero-energy states (each pair consists of an excitation and its Hermitian conjugated counterpart), while there is the only one pair of states with zero energy for the single-pair modes. The found zero-energy vortex modes are stable in a wide interval of values of the chemical potential, spin-orbit interaction parameter, and amplitudes of superconducting pairings. Nevertheless, these modes becomes gapped by changing the other model parameters, such as the difference of on-site energies for different orbitals, the hopping parameter between next-nearest-neighbor sites, and the difference of the hopping parameters between nearest sites in $x$ and $y$ directions of the square lattice.

The paper has been organized in five sections. In Sec. II
we describe conditions of the formation of the higher-order topological superconducting phase in the considered model. The topological invariant and relevant topological phase diagram are presented. In Sec. III we derive the analytical solution for zero-energy vortex-localized states in the model and explain the difference of the obtained solutions from the ones in the case of FOTSCs. The conditions of the coexistence of zero-energy vortex-localized modes and Majorana corner states are discussed in Sec. IV. We also explain in Sec. IV how zero-energy vortex modes becomes gapped. Sec. V is a summary.

\section{\label{sec2} Higher-order topological superconducting phase}
To investigate the possibility of vortex zero modes appearance in HOTSC we consider the two-level Hamiltonian with spin-orbit interaction and singlet superconducting coupling \cite{wangQ-18}
\begin{eqnarray} \label{Ham}
    &&H =\sum_{f\eta\sigma}\left(\eta\Delta\varepsilon-\mu\right)c^{\dag}_{f\eta\sigma}c_{f\eta\sigma}\\
    &&+\sum_{\eta}\eta\left( \sum_{\langle fm\rangle_{x},\sigma}t_{x}+\sum_{\langle fm\rangle_{y},\sigma}t_{y}+\sum_{\langle\langle fm\rangle\rangle,\sigma}t_{1}\right)c^{\dag}_{f\eta\sigma}c_{m\eta\sigma}\nonumber\\
    &&+i\lambda\sum_{\langle fm\rangle}\left[\hat{\tau}^{\alpha\beta},e_{fm}\right]_{z}\hat{\sigma}^{\nu\eta}_{x}c^{\dag}_{f\nu\alpha}c_{m\eta\beta}  + \nonumber
\end{eqnarray}
\begin{eqnarray}    
    &&+\left(\Delta_{x}\sum_{\langle fm\rangle_x,\eta} + \Delta_{y}\sum_{\langle fm\rangle_y,\eta}\right)c^{\dag}_{f\eta\uparrow}c^{\dag}_{m\eta\downarrow}\nonumber\\
&&+\Delta_{0}\sum_{f\eta}c^{\dag}_{f\eta\uparrow}c^{\dag}_{f\eta\downarrow}+\text{H.c.},\nonumber
\end{eqnarray}
where $c_{f\eta\sigma}$ annihilates an electron with a spin $\sigma$ on an $A,B$ orbital (with $\eta=\pm1$ correspondingly) at a square lattice site $f=\left(i,j\right)$; $i,j=1,...,N$; $\Delta\varepsilon$ is an on-site energy shift opposite for different orbitals; $\mu$ is a chemical potential. The intraorbital nearest-neighbor $t_{x,y}$ as well as next-nearest-neighbor $t_{1}$ hopping parameters are of opposite signs for different orbitals leading to the inverted bands. The parameter $\lambda$ defines an intensity of the interorbital Rashba spin-orbit coupling; $e_{fm}$ is a unit vector pointing along the direction of electron motion from the $m$th to $f$th site. The parameters $\Delta_{0,x,y}$ are intensities of the intraorbital on-site and intersite singlet pairing that results in overall $s_{\pm}$-wave superconductivity in the case $\Delta_x=\Delta_y$ or $s+d_{x^2-y^2}$-wave superconductivity in the case of $\Delta_x=-\Delta_y$. The Pauli matrices $\hat{\sigma}_{n}$ and $\hat{\tau}_{n}$ ($n=x,y,z$) act in the orbital and spin subspaces, respectively.

The elementary excitations of the Hamiltonian (\ref{Ham}) can be classified by spin projection number $\sigma$ and may be written in the form
\begin{eqnarray}
\label{selfexit}
\alpha_{\sigma}=\sum_{f}\left[u_{f\sigma}c_{fA\sigma}+v_{f\overline{\sigma}}c_{fB\overline{\sigma}}+w_{f\overline{\sigma}}c_{fA\overline{\sigma}}^{\dag}+z_{f\sigma}c_{fB\sigma}^{\dag}\right].
\nonumber \\
\end{eqnarray}

The bulk spectrum of the system in superconducting and nonsuperconducting regime is defined as
\begin{eqnarray}
\label{E_k}
&&\varepsilon_{sc}=\sqrt{(\varepsilon_{nosc}\pm\mu)^2+|\Delta_k|^2},\\
&&\varepsilon_{nosc}=\sqrt{t_k^2+|\lambda_k|^2},\nonumber\\
&&t_k=\Delta\varepsilon+2t_x\cos k_x + 2t_y\cos k_y + 4t_1\cos k_x\cos k_y,\nonumber\\
&&\Delta_k=\Delta_0+2\Delta_x\cos k_x+2\Delta_y\cos k_y,\nonumber\\
&&\lambda_{k\sigma}=2\lambda\left(\sin k_y-i\sigma\sin k_x\right)\nonumber,
\end{eqnarray}

This model is known to provide higher-order topological phase. In the case of $\sign(\Delta_x\Delta_y)=-\sign(t_x t_y)$ and $t_1\neq0$ ($|t_1| < (|t_x| + |t_y|)/2$) or $\Delta\varepsilon\neq0$ this model supports Majorana corner states, induced by the change of the effective mass of edge states at the corners \cite{wangQ-18,aksenov-23, aksenov-23-2}. The Hamiltonian (\ref{Ham}) supports time-reversal symmetry. Therefore, there is a pair of Majorana corner modes (Majorana Kramers pair) protected by TRS at each corner~\cite{wangQ-18}.

In the case of $t_1=\Delta\varepsilon=0$ (as it will be discussed in the next section, zero-energy vortex modes can exist in such case) the procedure of defining the effective mass of edge states, carried out previously, is not available, as the nonsuperconducting  bulk spectrum is not gapped any more. On the other hand, it is obviously that HOTSC phase must be preserved, since the bulk in \eqref{E_k} and edge superconducting spectrum remain gapped at $t_1=\Delta\varepsilon=0$ (see \cite{aksenov-23, aksenov-23-2}). To prove the presence of HOTSC phase in this case one can calculate the polarization of the Wannier bands and the quadrupole (the details of the procedure can be found in \cite{benalcazar-17PRB} and supplementary materials of \cite{Pahomi2020}), which are the bulk topological indexes and cannot be changed, until the superconducting bulk gap or Wannier bands gap closes. The numerical calculation in the case of $|t_x|=|t_y|$, $\Delta_x=-\sign(t_x t_y)\Delta_y$; $\mu,\Delta\varepsilon\rightarrow0$; $t_1,\Delta_0=0$ shows the Wannier band polarization to be $(p_x^{\nu_y^-},p_y^{\nu_x^-})=(1/2,1/2)$ along with quadrupole $q_{xy}=2p_x^{\nu_y^-}p_y^{\nu_x^-}=1/2$ corresponding to the HOTSC phase. At $|\Delta_0|>2|\Delta_x-\sign(t_x t_y)\Delta_y|$ these indexes become trivial $(p_x^{\nu_y^-},p_y^{\nu_x^-})=(0,0)$, $q_{xy}=0$ demonstrating the trivial superconducting phase.

\begin{figure}[ht]
\begin{center}
\includegraphics[width=0.35\textwidth]{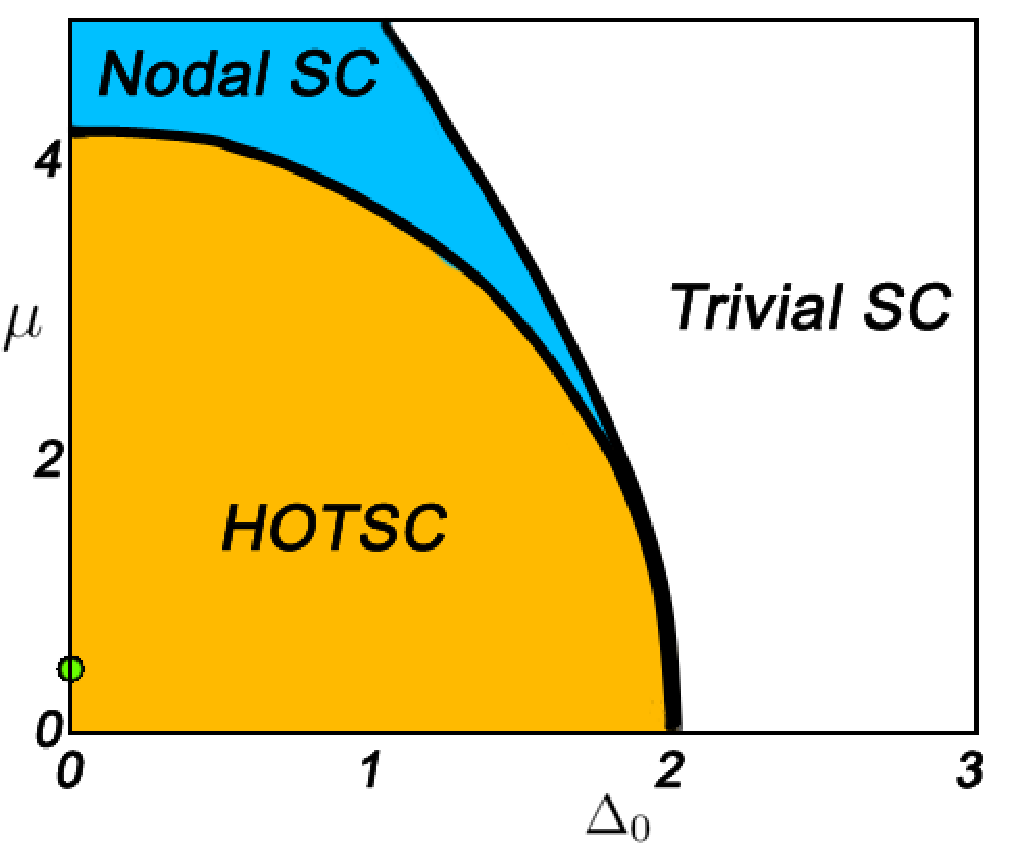}
\caption{The topological phase diagram in variables chemical potential $\mu$ - on-site superconducting coupling constant $\Delta_0$ for the 2D second-order topological superconductor (\ref{Ham}). Parameters are $\Delta\varepsilon=t_1=0$, $t_x=-t_y=2$, $\Delta_x=\Delta_y=0.5$, $\lambda=1.5$.}\label{concave_fig2}
\end{center}
\end{figure}

All the parameter space of the system, which is not separated from the specified nontrivial point by the mentioned above gaps closure, corresponds to the HOTSC phase providing topological corner excitations. In general, there is a combination of two different mechanisms of HOTSC phase destruction. The first is defined by the superconducting coupling constants: if one starts from the noticed before nontrivial point and increases the $\Delta_0$ or changes the ratio $\Delta_x/\Delta_y$, then one of the gaps (bulk or Wannier bands gap depending on the specific parameters) will close and reopen at critical value (for $\Delta\varepsilon=\mu=t_1=0$ it is $|\Delta_0|=2|\Delta_x-\sign(t_x t_y)\Delta_y|$) and the system turns into a trivial superconductor. On the other hand,  increasing $\mu$ leads to the crossing of the Fermi contour and $\Delta_k$ nodal lines turning the system into the nodal phase (for the parameters taken at the noticed above nontrivial point it happens at $|\mu|=2\sqrt{2}|\lambda|$). The exemplifying phase diagram for the investigated system, obtained with numerical calculations of the bulk band gap and topological indexes is depicted on the Fig.1.

\section{\label{sec3} Zero-energy vortex modes}

To introduce the superconducting vortex we use the modification of the superconducting coupling parameters with the notation:
\begin{eqnarray}
\label{Vortex}
&&\Delta_{fm}({\bf r})=\Delta_{fm}\exp\left\{il \arg\left[z({\bf R}_f+{\bf R}_m)/2-z({\bf R}_v)\right]\right\} \nonumber \\ 
&&~~~~~~~~\times \tanh\left(\frac{|{\bf R}_{f}+{\bf R}_{m}-2{\bf R}_v|}{2\xi}\right),
\end{eqnarray}
where $z({\bf R})=x+iy$, $\xi$ is coherence length, $l=\pm1$ for vortex/antivortex, and the vortex is set at the position ${\bf R}_v$. While we use this notation for numerical calculations, the analytical approach demonstrates that the precise form of the superconducting coupling modification does not matter as long, as it is centrally symmetric vortex-like.

We start our investigation of the zero-energy vortex modes with the case of $\Delta\varepsilon=t_1=0$ along with $|t_x|=|t_y|=t$. This case corresponds to the gapless bulk energy spectrum $\varepsilon_{nosc}$ in (\ref{E_k}) in the absence of superconducting coupling with gap closure at the pair of high-symmetry points: $(0,0)$, $(\pi,\pi)$ or $(0,\pi)$, $(\pi,0)$ for $\tau=\sign(t_x t_y)=\mp1$ correspondingly, which are the nodal points of the spin-orbit interaction.

To obtain the solution for zero-energy vortex modes at the chosen parameters we use the wide-known procedure \cite{fu-08, sau-10, fukui-10, rakhmanov-11, akzyanov-14,kobayashi-20} with one difference: the investigated system is two-level in our case. We make the expansion of the energy terms in Eq. (\ref{E_k}) in the vicinity of the spin-orbit interaction nodal points
\begin{eqnarray}
\label{tk_exp}
&&t_k\approx t\sign(t_x)c_x(k_y^2-k_x^2)\\
&&\lambda_{k\sigma}\approx -2\lambda c_x(\tau k_y+i\sigma k_x),\nonumber\\
&&\Delta_k\approx\Delta_0+2c_x(\Delta_x-\Delta_y\tau).
\end{eqnarray}
with $c_x=\cos k_x^{(N)}=\pm1$ referring to different nodal points. Obviously, that $\tau = -c_xc_y$ with $c_y = \cos k_y^{(N)}$.

As the spectrum is linear in quasi-momentum ${\bf k}$ around the nodal points we as a first step neglect the quadratic terms in $t_k$ and $\Delta_k$. We use the continuum approximation $(k_x,k_y)\rightarrow-i(\partial_x,\partial_y)$ (with distances, measured in the parameter of the unit cell), and include the modification of superconducting coupling to describe the vortex
\begin{eqnarray}
\label{sc_form}
\Delta_k\rightarrow\chi\Delta(r)\exp(il\phi),~\Delta(r\rightarrow\infty)>0,~\chi=\pm1.
\end{eqnarray}
Here $(r,\phi)$ are polar coordinates with the origin at the center of the vortex, and we separate out the sign of the coupling constant at the nodal point by introduction of $\chi$. As we are interested in zero-energy modes the final equations on coefficients (\ref{selfexit}) take the form
\begin{eqnarray}
\label{Ham_r}
&&\left[\begin{array}{cccc}
-\mu & \lambda_{r\sigma} & \sigma\Delta_{r\phi}^* & 0 \\
-\lambda_{r\sigma}^* & -\mu & 0 & -\sigma\Delta_{r\phi}^*\\
\sigma\Delta_{r\phi} & 0 & \mu & \lambda_{r\sigma}\\
0 & -\sigma\Delta_{r\phi} & -\lambda_{r\sigma}^* & \mu\end{array}\right]\left[\begin{array}{c}u_{\sigma}\\v_{\overline{\sigma}}\\w_{\overline{\sigma}}\\z_{\sigma}\end{array}\right]=0,\nonumber\\
&&\lambda_{r\sigma}=2\lambda\sigma c_x\exp(-i\sigma\tau\phi)\left(\partial_r-\frac{i\sigma\tau}{r}\partial_{\phi}\right),\\
&&\Delta_{r\phi}=\chi\Delta(r)\exp(il\phi).\nonumber
\end{eqnarray}

As it can be easily checked, these equations separate at $\mu=0$ into two pairs, which are equivalent up to conjugation along with replacement $\sigma\rightarrow\overline{\sigma}$. It means that if the $\alpha_{\sigma}$ is a zero-energy solution, then $\alpha_{\sigma}^{\dag}$ is also a solution.

The differential equation for $u_{\sigma}(r,\phi)$, $z_{\sigma}(r,\phi)$ parameters can be written in the next form
\begin{eqnarray}
\label{ZVM_eq1}
&&\left[\left(\partial_r^2-\frac{i\sigma\tau}{r^2}+\frac{1}{r^2}\partial_{\phi}^2\right)-\left(\frac{\Delta(r)}{2\lambda}\right)^2-\right.\\
&&-\left.\left(\frac{d\Delta(r)/dr}{\Delta(r)}-\frac{1+l\sigma\tau}{r}\right)\left(\partial_r+\frac{i\sigma\tau}{r}\partial_{\phi}\right)\right]u_{\sigma}=0,\nonumber\\
&&z_{\sigma}=-2\lambda\chi c_x\Delta(r)\exp(i\phi(l+\sigma\tau))\left(\partial_r+\frac{i\sigma\tau}{r}\partial_{\phi}\right)u_{\sigma}.\nonumber
\end{eqnarray}

This equation can be solved in the case of $\sigma=-\tau l$ as an exponentially decreasing (increasing) function corresponding to the vortex-localized (boundary-localized) excitation. However, these equations must be accompanied with boundary equations to investigate the fate of the increasing solution. Actually, the system of equations has been extensively studied for FOTSCs~\cite{fu-08, rakhmanov-11, akzyanov-14, akzyanov-16, kobayashi-20}. The difference in our case is as follows: usually it is supposed for FOTSCs, that exponentially increasing solutions correspond to an edge state, localized on the boundary (with the exception of \cite{akzyanov-16}, where it is localized at finite radius). In HOTSC the edge states are gapped and there is no such a zero mode counterpart. Therefore, there is only a pair of vortex-localized zero-energy excitations in \eqref{ZVM_eq1} in our case. This solution has the form
\begin{eqnarray}
\label{ZVM_S1}
&&\left[\begin{array}{c} u_{\sigma} \\ z_{\sigma}\end{array}\right]=\frac{1}{\sqrt{\mathcal{N}}}\left[\begin{array}{c} 1 \\ s\end{array}\right]F(r),~~\sigma=-\tau l,\\
&&F(r)=\exp\left(-\int\limits_0^r d\rho \frac{\Delta(\rho)}{2|\lambda|}\right),~~s=c_x\chi\sign\lambda,\nonumber
\end{eqnarray}
accompanied with a pair of their hermitian conjugated counterparts (solutions for $w_{\sigma}$ and $v_\sigma$).
It is seen that the double-pair zero-energy vortex modes appear due to two distinct nodal points with $c_x=\pm1$. Previously, the possibility of formation of multiply vortex zero modes was discussed in~\cite{hosur-11, kobayashi-20} for FOTSCs.

\begin{figure*}[ht]
\begin{center}
\includegraphics[width=0.8\textwidth]{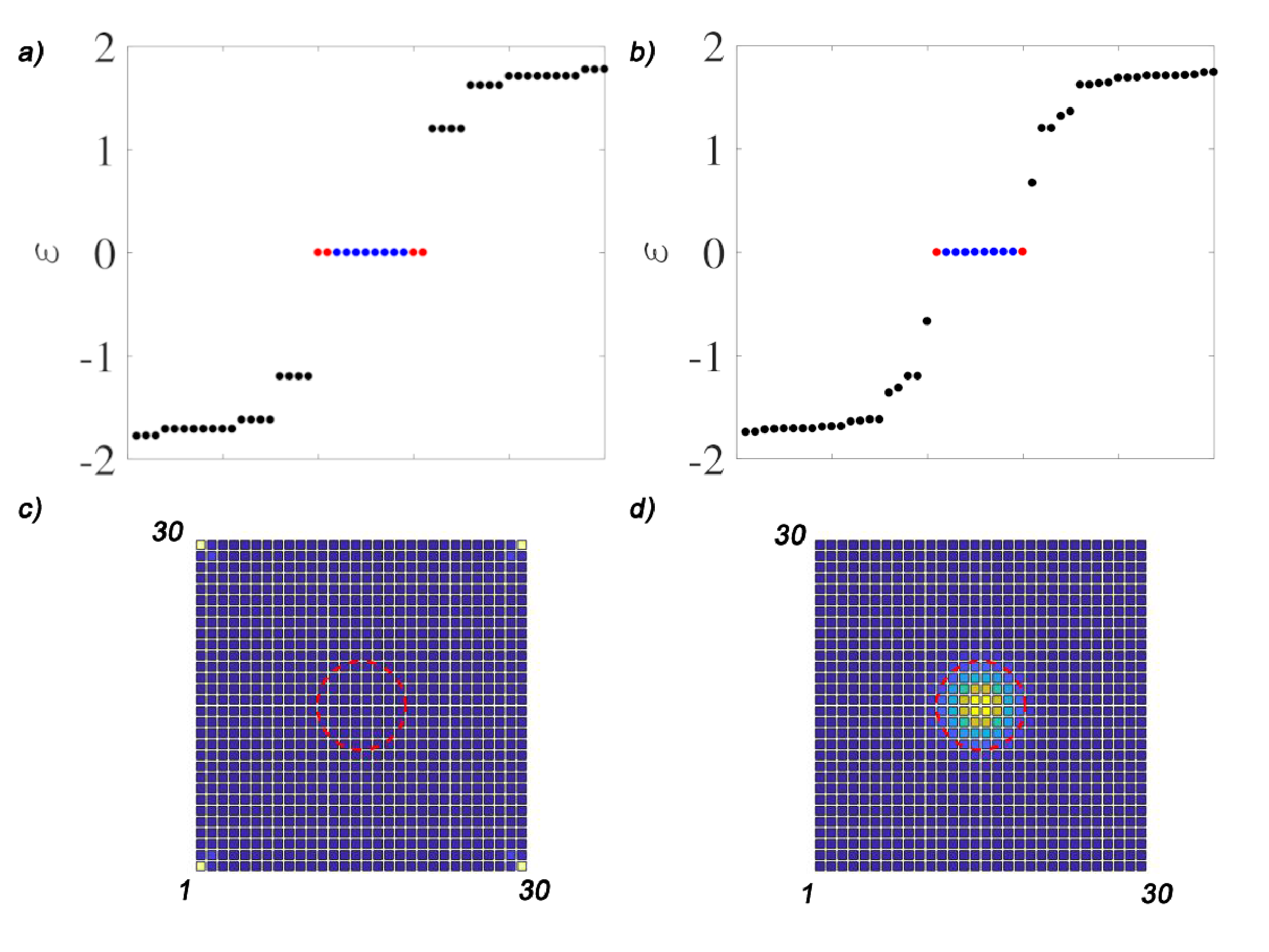}
\caption{a) The energy spectrum of the square-shaped system at $\Delta\varepsilon=t_1=0$, $t_x=-t_y=2$, $\Delta_x=\Delta_y=0.5$, $\lambda=1.5$, $\mu=0.5$, $\Delta_0=0$ (green point in Fig. 1) with a vortex located at the center of the system and characterizing by $l=1$, $\xi=4$. Blue dots correspond to the topological corner modes, red dots are for the vortex zero modes. b) The energy spectrum corresponding to formation of the single-pair vortex mode instead of double-pair vortex modes as in a). Parameters are $t_x = 2$, $t_y = -1.8$; $t_1=-0.05$ and $\Delta\varepsilon=-0.2$ are obtained from the relation \eqref{ZVM_cond}, the other parameters are the same as in a). c),d) The corresponding spatial distributions of corner and vortex excitations for the parameters in a), red dashed line depicts the circle of radius $\xi$ around the vortex center.}\label{concave_fig}
\end{center}
\end{figure*}

One can notice that there is also a mathematical solution with $\sigma=\tau l$.
\begin{eqnarray}
\label{ZVM_S2}
&&\left[\begin{array}{c} u_{\sigma} \\ z_{\sigma}\end{array}\right]=\frac{1}{\sqrt{\mathcal{N}}}\left[\begin{array}{c} \exp(-il\phi) \\ s\exp(il\phi)\end{array}\right]\frac{F(r)}{r}.
\end{eqnarray}
While it has no physical sense at $\mu=0$, it represents a necessary additive to obtain a solution for $\mu\neq0$. Making the substitution of the combination of the first solution (\ref{ZVM_S1}) with a factor $f(r)$ for $u_{\sigma}$, $z_{\sigma}$ and the second solution (\ref{ZVM_S2}) with a factor $g(r)$ for $v_{\overline{\sigma}}$, $w_{\overline{\sigma}}$ (see Appendix \ref{apx_ZVM} for details) one can obtain the final form
\begin{eqnarray}
\label{ZVM_S3}
&&\left[\begin{array}{c} u_{\sigma} \\ v_{\overline{\sigma}} \\ w_{\overline{\sigma}} \\ z_{\sigma}\end{array}\right]=
\frac{1}{\sqrt{\mathcal{N}}}\left[\begin{array}{c} J_0\left(r\mu/2\lambda\right) \\ J_1\left(r\mu/2\lambda\right)\exp(-il\phi) \\ sJ_1\left(r\mu/2\lambda\right)\exp(il\phi) \\ sJ_0\left(r\mu/2\lambda\right)\end{array}\right]F(r),
\end{eqnarray}
with $\sigma=-l\tau$. It can be concluded that there are zero-energy vortex modes only with one spin projection $\sigma$ defined by the vortex vorticity sign $l$. So, at finite $\mu$ the system still provides two zero-energy vortex-localized excitations and their two Hermitian conjugated counterparts, corresponding to two nodal points, at which the non-superconducting bulk energy spectrum (\ref{E_k}) is gapless.

Obtained zero-energy vortex modes are not protected by any topological number. Despite that, according to (\ref{ZVM_S3}) their existence at $\Delta\varepsilon=t_1=0$, $|t_x|=|t_y|$ is insensitive to the precise values of the spin-orbit interaction $\lambda$, chemical potential $\mu$, and superconducting coupling parameters $\Delta_{0/x/y}$ (apart from the case $\Delta (k_x^{(N)},k_y^{(N)}) = 0$). This result is also confirmed by the numerical calculations. Moreover, as it will be shown in the next paragraph, the realization of gapless vortex modes is not restricted by the discussed parametric point.

\section{\label{sec4} Coexistence of zero-energy vortex and topological corner modes}

Now we return to the generalized form of $t_k$ in \eqref{E_k} with unrestricted parameters $t_x$, $t_y$, $\Delta\varepsilon$, and $t_1$, and make the expansion up to quadratic terms in $k_{x,y}$
\begin{eqnarray}
\label{tk_exp2}
&&t_k\approx\widetilde{\varepsilon}-tc_x\sign t_x(k_x^2-k_y^2)-\\
&&-\sign t_x(c_x\delta t-2t_1\sign t_y)(k_x^2+k_y^2),\nonumber\\
&&\widetilde{\varepsilon}=\Delta\varepsilon+2\sign t_x(2c_x\delta t-2t_1\sign(t_y)),\nonumber\\
&&t=(|t_x|+|t_y|)/2,~~\delta t=(|t_x|-|t_y|)/2.\nonumber
\end{eqnarray}

The first term $\widetilde{\varepsilon}$ in $t_k$ (\ref{tk_exp2}) opens a gap in the spectrum $\varepsilon_{nosc}$ at a nodal point and gaps out the vortex modes (\ref{ZVM_S3}). It is easy to see that size of the gap of the vortex modes  $\sim\widetilde{\varepsilon}$. The second term in the continuum limit contains a combination of differential operators, 
which does not influence on the zero-energy solution (\ref{ZVM_S3}) due to its specific dependence on $\phi$, justifying neglecting this term previously in Eqs \eqref{Ham_r} (see Appendix \ref{apx_kinEn}).

Meanwhile, the effect of the third term is nonzero. Consequently, the vortex modes remain gapless, if both first and third terms in (\ref{tk_exp2}) are zero corresponding to the existence of Dirac cones in the non-superconducting bulk spectrum and the absence of the additive to the vortex modes energy from the kinetic term. This leads to the condition
\begin{eqnarray}
\label{ZVM_cond}
&&t_1=c_x\delta t\sign t_y/2,\\
&&\Delta\varepsilon=-2c_x \delta t \sign t_x.\nonumber
\end{eqnarray}
It is seen that for $|t_x| = |t_y|$, $\delta t = 0$ we get  $\Delta\varepsilon = t_1 = 0$, the case of which is considered in the previous section. 

The parameter ratio (\ref{ZVM_cond}) represents the condition of appearance of the zero-energy vortex modes in the system. To obtain coexistence of Majorana corner modes and zero vortex modes this condition must be satisfied with the conditions of HOTSC phase formation. As it was mentioned in Sec. \ref{sec2}, the necessary condition for the topological corner modes to be realized is $\sign(\Delta_x\Delta_y)=-\sign(t_x t_y) \equiv -\tau$. It should be noted that such restriction is absent for the vortex zero modes. Therefore, if the equality \eqref{ZVM_cond} along with $\sign(\Delta_x\Delta_y)=-\tau$ are satisfied, then the coexistence of HOTSC state and zero-energy vortex modes is characterized by the following conditions for the other parameters: 1) chemical potential $\mu$ can not lie in the nodal phase (see Fig. 1), where the bulk spectrum $\varepsilon_{sc}$ in \eqref{E_k} is gapless and corner, as well as vortex excitations are mixed with the bulk ones (this interval of $\mu$ is mainly determined by the spin-orbit coupling parameter $\lambda$); 2) we can increase superconducting coupling constant $\Delta_0$ from zero value or change the ratio $\Delta_x/\Delta_y$ starting from the point $-\tau$ until the bulk or Wannier bands gaps are not close (after reopening one of these gaps the topological corner excitations disappear, as it is seen from the topological phase diagram in Fig. 1, while the vortex zero modes still exist).

The energy spectrum of the square-shaped system with parameters $\Delta\varepsilon=t_1=0$, $t_x=-t_y=2$, $\Delta_x=\Delta_y=0.5$, $\lambda=1.5$, $\Delta_0 = 0$, $\mu = 0.5$ is shown in Fig. 2(a). The system supports both topological corner-localized excitations (Fig. 2(c)) and zero-energy vortex-localized excitations (Fig. 2(d)). In the presented case both pairs of vortex localized modes have the same localization length. Meanwhile, it should be noted that in the case of $\Delta_0\neq0$ the wave functions of the vortex-localized modes become different as they depend on the value of $|\Delta_k|$, which is different at two Dirac points for $\Delta_0\neq0$.

\begin{figure*}[ht]
\begin{center}
\includegraphics[width=0.9\textwidth]{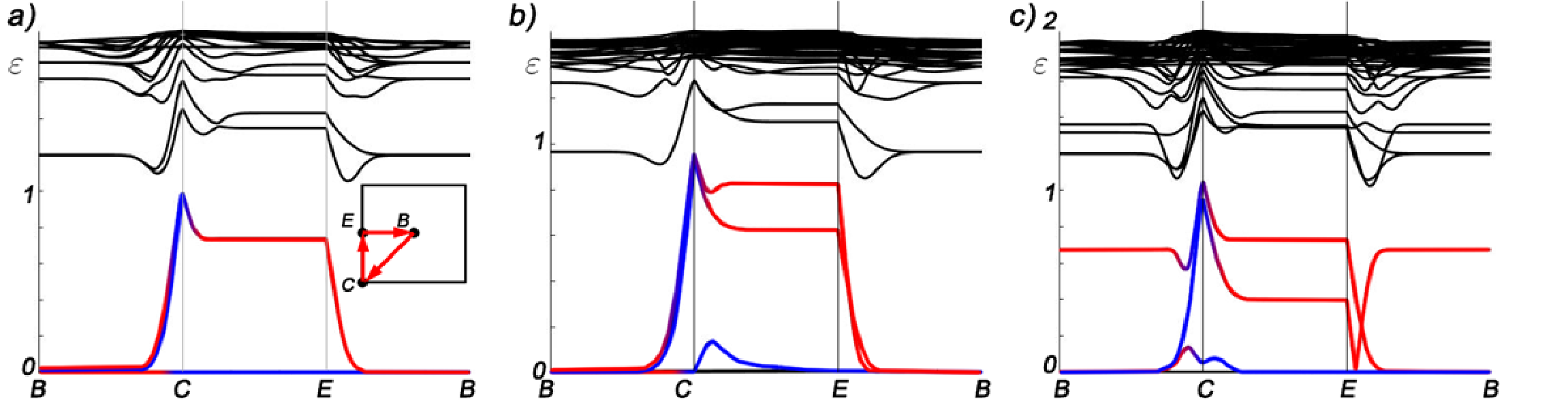}
\caption{Dependence of excitation energy of the square-shaped 2D topological superconductor on the vortex position, which moves through the closed path: bulk center (B) $\rightarrow$ corner (C) $\rightarrow$ edge center (E) $\rightarrow$ bulk center. Red line is for vortex excitations, blue line for corner excitation at the perturbed corner C. a) The case of double-pair zero-energy vortex-localized excitations with $\Delta_0=0$. b) The same as (a) but with $\Delta_0=0.5$. c) The case of the single-pair zero-energy vortex modes with $\Delta_0=0$. }\label{concave_fig3}
\end{center}
\end{figure*}

In the case of $|t_x|\neq|t_y|$, $\delta t \ne 0$ and satisfying the relations \eqref{ZVM_cond} the bulk non-superconducting energy spectrum $\varepsilon_{nosc}$ in \eqref{E_k} is gapless only at one nodal point, while the gap at the second nodal point remains finite. Therefore, only single-pair vortex modes with zero energy can exist at $|t_x|\neq|t_y|$. The energy spectrum for this case is presented in Fig. 2(b). The parameters are $\delta t = 0.1$, $\tau = -1$, $\Delta_x=\Delta_y=0.5$, $\lambda=1.5$, $\Delta_0 = 0$, $\mu = 0.5$, $t_1$, as well as $\Delta\varepsilon$, are determined from the relations \eqref{ZVM_cond}. Although there is only single-pair vortex-localized gapless modes for such parameters, their wave functions remain the same as in the case of a double-pair vortex-localized modes. This is related to the fact that the wave functions in both cases are defined by the same equations~\eqref{Ham_r}.

It can be concluded, that while the existence of Dirac cones is a necessary but not sufficient condition for the realization of zero-energy vortex-localized excitations, the number of these excitations, if they appear, is defined by the number of Dirac cones. It should be noted that the correspondence between the zero-energy solutions in the system with fermion-vortex interaction and Dirac cones were reported before in several studies starting with \cite{jackiw-81}. 

To examine the interaction between corner and vortex zero modes we calculate the spectrum of the finite square-shaped system with different positions of the vortex core: in the bulk, near the edge and near one of the corners. The interaction appears to be rather complicated and dependent on the kinetic and superconducting parameters of the model. Here we consider only the case of coexistence of corner and vortex-localized zero modes.

Approaching of the vortex core along the diagonal to the corner on the distance of $\sim\xi$ (path BC in Fig.3(a)) gaps out the initial corner mode localized at corner C.  If $\delta t=0$ (the case of a double-pair zero-energy vortex modes) one pair of the vortex-localized modes remains gapless, while the other becomes gapped (Fig.3(a,b)). In close proximity of the vortex core to the corner all excitations under consideration including the gapless vortex modes become corner-localized (while at this moment vortex-localized and corner localized modes can not be clearly distinguished). Proceeding further along path CE shown in Fig.3(a), when the vortex core moves along the edge from the corner, the gapped excitations transform to vortex-localized, while the gapless excitation remains corner-localized. In the case of $\Delta_0=0$ the zero mode remains gapless (Fig.3(a)), otherwise it becomes slightly gapped (Fig.3(b)), until the vortex core moves far away from the corner.

In the case of $\delta t\neq0$ both the corner excitation and the single-pair vortex-localized modes become gapped (but the last one with a sufficiently smaller gap) when the vortex core approaches the corner along path BC (Fig.3(c)). Similarly to the previous case, the vortex-localized excitation transforms to the corner-localized as the vortex core getting closer to the corner, and becomes gapless, when the core moves far away along the edge CE.

The variety of the results, which becomes broader when we consider the cases of vortex-localized zero modes without HOTSC phase or HOTSC phase with gapped vortex modes, is due to several reasons. Firstly, the vortex modes are not topologically protected and can be gapped out by sufficient defects, such as the system boundary. Secondly, the corner excitation and vortex excitation are gapped due to their hybridization (we appreciate the referee for focusing our attention to this possibility). Thirdly, the vortex core itself (without zero-energy modes) breaks the symmetry, which protects Majorana Kramers pair, making them unstable and can lead to their hybridization, when the vortex core is near a corner. The precise combination of these factors can be different depending on the parameters and can lead to an essentially different behavior of the considered excitations, when the vortex is located near the corner.

\section{Summary}

We investigated the possibility of appearance of zero-energy vortex-localized excitations in the 2D second-order topological superconductor. To obtain the result we used the two-orbital model, which is known to provide Majorana corner states, with spin-orbit interaction between different orbitals and superconducting coupling with the enhanced $s$-wave or $s+d_{x^2-y^2}$-wave symmetries. 

We analytically demonstrated, that if the bulk energy spectrum in the absence of superconducting coupling is gapless and has Dirac cones, then at specific kinetic parameters ratio there appear one or two pairs of zero-energy vortex-localized modes in the presence of superconductivity, according to the number of the initial Dirac points. The analytical solution for zero-energy vortex modes in the second-order topological superconductor was derived. The vortex excitations have only one spin projection defined by the vortex vorticity $l = \pm 1$. Contrary to the FOTSCs, there is no boundary-localized zero-energy counterpart for the vortex-localized modes, as the edge spectrum is gapped. It was shown that the zero-energy vortex modes can be gapped considering nonzero values of the model parameters, such as the difference of on-site energies for different orbitals, the hopping parameter between next-nearest-neighbor sites, or the difference of the hopping parameters between nearest sites in $x$ and $y$ directions of the square lattice. Therefore, in general, the zero-energy vortex modes can be unstable in HOTSC and this instability should be kept in mind for an experimental study of vortex modes in a system which is supposed to be HOTSC.

It was shown that different regimes can be realized in the parameter space of the considered model: HOTSC phase with the only Majorana corner modes, the region with the only zero-energy vortex modes (double-pair or single-pair), the region of coexistence of such vortex-localized and corner-localized excitations, the nodal phase, in which bulk excitations have near zero energy, and the trivial phase where any localized modes with zero energy are absent. The analytically obtained results were confirmed by the numerical study of the finite-size lattice. The coexistence of the Majorana corner modes and zero-energy vortex excitations was shown for the predicted parameters in the case of the vortex core locating away from the boundaries. If the vortex core approaches the boundary, then the vortex-localized excitations become gapped due to their interplay with the boundary modes. If the vortex core approaches the corner, the initial corner excitation becomes gapped and one pair of the vortex-localized modes transforms into the corner-like. Depending on the specific parameters of the model the low-energy modes can be gapless or slightly gapped when the vortex is located near the corner.

\begin{acknowledgments}
The authors acknowledge fruitful discussions with S.V. Aksenov.
The reported study was supported by Russian Science Foundation (project No. 24-22-20088, \href{https://rscf.ru/en/project/24-22-20088/}{https://rscf.ru/en/project/24-22-20088/}) and Krasnoyarsk Regional Fund of Science.
\end{acknowledgments}

\appendix

\section{\label{apx_ZVM} Zero-energy vortex modes at finite chemical potential}

The combination of solutions with allowed $r$-dependent coefficients for $\sigma= -\tau l$ may be written in the form
\begin{eqnarray}
\label{ZVM_A1}
&&\left[\begin{array}{c} u_{\sigma} \\ v_{\overline{\sigma}} \\ w_{\overline{\sigma}} \\ z_{\sigma}\end{array}\right]=
\frac{1}{\sqrt{\mathcal{N}}}\left[\begin{array}{c} f(r) \\ g(r)\exp(-il\phi+i\theta) \\ sg(r)\exp(il\phi+i\theta) \\ sf(r)\end{array}\right]F(r),
\end{eqnarray}
Substitution this anzats to (\ref{ZVM_eq1}) reduces them to a pair of equations
\begin{eqnarray}
\label{ZVM_eq2}
&&\partial_rf(r)=\beta g(r),~~ \beta=\frac{\mu}{2|\lambda|}\tau l c_x\exp(i\theta),\nonumber\\
&&\partial_r(rg(r))=-\beta rf(r).
\end{eqnarray}
In the case $\mu=0$ this leads to $f(r)=const$, $g(r)=const/r$. Otherwise this pair of equations reduces to the recurrent relations for Bessel functions
\begin{eqnarray}
\label{Bessel_rel}
\left(\frac{1}{x}\frac{d}{dx}\right)^m\left[x^nJ_n(x)\right]=x^{n-m}J_{n-m}(x),
\end{eqnarray}
with $x=\beta r$, $n=0,1$, $m=1$, if $\beta$ is real, thus leading to the final form of the zero-energy excitation (\ref{ZVM_S3}).

\section{\label{apx_kinEn} Kinetic operators extension}

The second term of (\ref{tk_exp2}) in the continuum limit contains the differential operator combination
\begin{eqnarray}
\label{tk_exp3}
&&(\partial_x^2-\partial_y^2)=\cos2\phi\left(\partial_r^2-\frac{\partial_r}{r}-\frac{\partial_{\phi}^2}{r^2}\right)+\\
&&~~~~~~~~~~~~~~+2\frac{\sin2\phi}{r}\left(\frac{\partial_{\phi}}{r}-\partial_r\partial_{\phi}\right).\nonumber
\end{eqnarray}
Obviously, due to specific $\phi$ dependence its effect on zero-energy vortex solutions (\ref{ZVM_S3}) $\Psi^{\dag}(\partial_x^2-\partial_y^2)\Psi=0$. Meanwhile the third term
\begin{eqnarray}
\label{tk_exp4}
(\partial_x^2+\partial_y^2)=\partial_r^2+\frac{\partial_{\phi}}{r}+\frac{\partial_{\phi}^2}{r^2},
\end{eqnarray}
has no such dependence, thus leading to the finite gap, if it is present in the Hamiltonian.

\bibliography{vort_corner_241122}

\end{document}